\documentclass[12pt]{elsart}
\usepackage{graphicx,cite}
\usepackage{amssymb}
\usepackage{upgreek}
\usepackage{color}
\journal{Physics Letters B}

\newcommand{\affuni}[2]{Dipartimento di Fisica dell'Universit\`a #1, #2, Italy.}
\newcommand{\affinfn}[2]{INFN Sezione di #1, #2, Italy.}

\def\ifm#1{\relax\ifmmode#1\else$#1$\fi}  \def\to{\ifm{\rightarrow}}  \def\epm{\ifm{e^+e^-}}
\def\gam{\ifm{\gamma}}     \def\x{\ifm{\times}}  \def\ab{\ifm{\sim}}    \def\dif{\hbox{d}}
\def\up#1;{$^{#1}$}  \def\dn#1;{$_{#1}$}    \def\DAF{DA\char8NE}  \def\f{\ifm{\phi}}
\def\Li{\ifm{L_i}}  \def\pic{\ifm{\pi^+\pi^-}}  \def\po{\ifm{\pi^0}}
\def\pt#1;#2;{\ifm{#1\x10^{#2}}}  \def\deg{\ifm{^\circ}}
\def\mt{\ifm{M_{\rm trk}}}

\begin{document}
\begin{frontmatter}
\title{\mathversion{bold} Search for light vector boson production in $e^+e^- \rightarrow \mu^+ \mu^- \gamma$ interactions with the KLOE experiment$^\star$\protect\thanks{This paper is dedicated to the memory of Juliet Lee-Franzini.}}
\collab{The KLOE-2 Collaboration}
\author[Frascati]{D.~Babusci},
\author[Cracow]{I.~Balwierz-Pytko},
\author[Frascati]{G.~Bencivenni},
\author[Frascati]{C.~Bloise},
\author[Frascati]{F.~Bossi},
\author[INFNRoma3]{P.~Branchini},
\author[Roma3,INFNRoma3]{A.~Budano},
\author[Uppsala]{L.~Caldeira~Balkest\aa hl},
\author[Roma3,INFNRoma3]{F.~Ceradini},
\author[Frascati]{P.~Ciambrone},
\author[Messina,INFNCatania]{F.~Curciarello\ead{fcurciarello@unime.it}},
\author[Cracow]{E.~Czerwi\'nski},
\author[Frascati]{E.~Dan\`e},
\author[Messina,INFNCatania]{V.~De~Leo},
\author[Frascati]{E.~De~Lucia},
\author[INFNBari]{G.~De~Robertis},
\author[Frascati]{A.~De~Santis},
\author[Frascati]{P.~De~Simone},
\author[Roma3,INFNRoma3]{A.~Di~Cicco},
\author[Roma1,INFNRoma1]{A.~Di~Domenico},
\author[INFNRoma2]{R.~Di~Salvo},
\author[Frascati]{D.~Domenici},
\author[Bari,INFNBari]{O.~Erriquez},
\author[Bari,INFNBari]{G.~Fanizzi},
\author[Roma2,INFNRoma2]{A.~Fantini},
\author[Frascati]{G.~Felici},
\author[ENEACasaccia,INFNRoma1]{S.~Fiore},
\author[Roma1,INFNRoma1]{P.~Franzini},
\author[Cracow]{A.~Gajos},
\author[Roma1,INFNRoma1]{P.~Gauzzi},
\author[Messina,INFNCatania]{G.~Giardina},
\author[Frascati]{S.~Giovannella},
\author[INFNRoma3]{E.~Graziani},
\author[Frascati]{F.~Happacher},
\author[Uppsala]{L.~Heijkenskj\"old}
\author[Uppsala]{B.~H\"oistad},
\author[Uppsala]{T.~Johansson},
\author[Cracow]{K.~Kacprzak},
\author[Cracow]{D.~Kami\'nska},
\author[Cracow]{W.~Krzemien},
\author[Uppsala]{A.~Kupsc},
\author[Frascati,StonyBrook]{J.~Lee-Franzini},
\author[INFNBari]{F.~Loddo},
\author[Roma3,INFNRoma3]{S.~Loffredo},
\author[Messina,INFNCatania,CentroCatania]{G.~Mandaglio\ead{gmandaglio@unime.it}},
\author[Moscow]{M.~Martemianov},
\author[Frascati,Marconi]{M.~Martini},
\author[Roma2,INFNRoma2]{M.~Mascolo},
\author[Roma2,INFNRoma2]{R.~Messi},
\author[Frascati]{S.~Miscetti},
\author[Frascati]{G.~Morello},
\author[INFNRoma2]{D.~Moricciani},
\author[Cracow]{P.~Moskal},
\author[INFNRoma3,LIP]{F.~Nguyen},
\author[Frascati]{A.~Palladino},
\author[INFNRoma3]{A.~Passeri},
\author[Energetica,Frascati]{V.~Patera},
\author[Roma3,INFNRoma3]{I.~Prado~Longhi},
\author[INFNBari]{A.~Ranieri},
\author[Frascati]{P.~Santangelo},
\author[Frascati]{I.~Sarra},
\author[Calabria,INFNCalabria]{M.~Schioppa},
\author[Frascati]{B.~Sciascia},
\author[Cracow]{M.~Silarski},
\author[INFNRoma3]{L.~Tortora},
\author[Frascati]{G.~Venanzoni\ead{graziano.venanzoni@lnf.infn.it}},
\author[Warsaw]{W.~Wi\'slicki},
\author[Uppsala]{M.~Wolke},
\author[Cracow]{J.~Zdebik}
\address[Bari]{\affuni{di Bari}{Bari}}
\address[INFNBari]{\affinfn{Bari}{Bari}}
\address[CentroCatania]{Centro Siciliano di Fisica Nucleare e Struttura della Materia, Catania, Italy.}
\address[INFNCatania]{\affinfn{Catania}{Catania}}
\address[Calabria]{\affuni{della Calabria}{Cosenza}}
\address[INFNCalabria]{INFN Gruppo collegato di Cosenza, Cosenza, Italy.}
\address[Cracow]{Institute of Physics, Jagiellonian University, Cracow, Poland.}
\address[Frascati]{Laboratori Nazionali di Frascati dell'INFN, Frascati, Italy.}
\address[Messina]{Dipartimento di Fisica e Scienze della Terra dell'Universit\`a di Messina, Messina, Italy.}\address[Moscow]{Institute for Theoretical and Experimental Physics (ITEP), Moscow, Russia.}
\address[Energetica]{Dipartimento di Scienze di Base ed Applicate per l'Ingegneria dell'Universit\`a
``Sapienza'', Roma, Italy.}
\address[Marconi]{Dipartimento di Scienze e Tecnologie applicate, Universit\`a ``Guglielmo Marconi", Roma, Italy.}
\address[Roma1]{\affuni{``Sapienza''}{Roma}}
\address[INFNRoma1]{\affinfn{Roma}{Roma}}
\address[Roma2]{\affuni{``Tor Vergata''}{Roma}}
\address[INFNRoma2]{\affinfn{Roma Tor Vergata}{Roma}}
\address[Roma3]{Dipartimento di Matematica e Fisica dell'Universit\`a
``Roma Tre'', Roma, Italy.}
\address[INFNRoma3]{\affinfn{Roma Tre}{Roma}}
\address[ENEACasaccia]{ENEA UTTMAT-IRR, Casaccia R.C., Roma, Italy}
\address[StonyBrook]{Physics Department, State University of New
York at Stony Brook, USA.}
\address[Uppsala]{Department of Physics and Astronomy, Uppsala University, Uppsala, Sweden.}
\address[Warsaw]{National Centre for Nuclear Research, Warsaw, Poland.}
\address[LIP]{Present Address: Laborat\'orio de Instrumenta\c{c}\~{a}o e F\'isica Experimental de Part\'iculas,
Lisbon, Portugal.}

\begin{abstract}
We have searched for a light vector boson $U$, the possible carrier of a ``dark force'', with the KLOE detector at the DA$\Phi$NE \epm\ collider, motivated by astrophysical evidence for the presence of dark matter in the universe. Using \epm\ collisions collected with an integrated luminosity of $239.3$~pb$^{-1}$, we look for a dimuon mass peak in the reaction \epm\to$\mu^+ \mu^-$\gam, corresponding to the decay $U\to\mu^+\mu^-$.
We find no evidence for a $U$ vector boson signal. We set a 90\%  CL upper limit for
the mixing parameter squared between the photon and the $U$ boson of 1.6$\x$10$^{-5}$ to 8.6$\x$10$^{-7}$ for the mass region $520<m_{\rm U}<980$ MeV.
\end{abstract}
\end{frontmatter}
\clearpage
\def\spreadlines#1{\par\renewcommand\baselinestretch{#1}\normalsize}
\def\red{\color{red}}  \def\blue{\color{blue}}

\section{Introduction}\label{Introduction}
 
Gravitational anomalies observed in large astronomical bodies are 
experimentally well established and are often interpreted as an excess in mass over the visible matter by more than a factor of five.
Dark matter (DM) is at present detected only by these
gravitational effects and in the cosmic microwave background and its nature remains as yet unknown. It is also well established that baryons can only contribute minutely to DM~\cite{PDG}. 
 There are several well motivated models in which DM consists
of new particles belonging to
a secluded gauge sector under which the SM particles are
uncharged~\cite{Holdom,U_th1,U_th2,Fayet,U_th6} . In the minimal
setup, the new interaction is mediated by a new gauge vector boson, 
the $U$ boson\footnote{Also referred to as A$^{\prime}$.},
which can kinetically mix with the ordinary photon through 
high-order diagrams, providing therefore a small coupling between the
$U$  and SM particles~\cite{Holdom,U_th1,U_th2,Fayet,U_th6}. This mechanism can be parametrized 
by a single mixing parameter, $\epsilon$, equal to the ratio of dark
and standard model electromagnetic couplings~\cite{Holdom}. Recently, the existence of a $U$ boson of mass $\mathcal{O}$(1GeV) and $\epsilon$ in the range 10\up-2;--10\up-7;, has been advocated
to explain several puzzling effects observed in astrophysics experiments, 
which fail standard astrophysical interpretations~\cite{U_th3,U_th4,U_th5,Pamela,AMS,Integral,Atic,Hess,Fermi,Dama/Libra,CDMS,COGENT,CRESST}.

High luminosity \epm\ colliders are ideal tools for the search of the $U$ boson~\cite{Essig,Batell,Reece,babayaga_article} because
 they provide a clean environment and good understanding of background.
A particularly clean channel is the reaction $e^+e^- \to U \gam$ followed by the decay $U\to\ell^+ \ell^- $, where $\ell= e,\ \mu$. Production of $U$ boson would result in a peak in the dilepton invariant mass spectrum.
Currently, the $U\gam$ production process allows one to reach a sensitivity of the mixing parameter $\epsilon$ in the range 10\up-3;--10\up-2;, for $U$-boson masses, $M_{\rm U}$,  up to a few GeV~\cite{U_th1,U_th2,Fayet,U_th6,babayaga_article}.

The search for a $U$-boson signal described in the following employs data collected in 2002 with the KLOE detector at \DAF, running at the \f-meson peak, with an integrated luminosity of 239.3 pb\up-1;. We limit our search to the muon channel, searching for a peak in the dimuon mass spectrum. The process \epm\to$\mu^+\mu^-$\gam\ receives a very large contribution from the reaction \epm\to$\mu^+\mu^-$ with additional photon emission by electrons or muons, usually called initial and final state radiation or ISR and FSR. Kinematical and geometrical cuts strongly suppress the FSR contribution. The ISR contribution can be written as
\begin{equation}
{\dif\sigma_{\mu\mu\gam}\over\dif M_{\mu\mu}}=\sigma(\epm\to\mu^+\mu^-, M_{\mu\mu})\cdot H\,  ,
\label{eq.1}
\end{equation}
where $\dif\sigma_{\mu\mu\gam}/\dif M_{\mu\mu}$ is the differential cross section
for $\epm\to\mu^+\mu^-\gamma$ as a function of the dimuon invariant mass $M_{\mu\mu}$, and $H$ is the radiator function. $H$ has been obtained from QED including NLO corrections~\cite{H,H_1,H_2,PHOKHARA,H_3}. Comparison with the measured cross section allows the extraction of a limit for $\epsilon$.

\section{The KLOE Detector}\label{KLOE}

The KLOE detector operates at DA$\Phi$NE, the Frascati $\phi$-factory. \DAF\ is an \epm\  collider usually operated at a center of mass energy, $W\sim m_\phi\ab1.019$ GeV. Positron and electron beams collide at an angle of $\pi-$25 mrad, producing \f\ mesons nearly at rest. The KLOE detector consists of a large cylindrical drift chamber (DC)~\cite{KLOE_DC}, surrounded by a lead scintillating-fiber electromagnetic calorimeter (EMC)~\cite{KLOE_EMC}.  
A superconducting coil around  the EMC provides a 0.52 T magnetic field along the bisector of the colliding beams. The bisector is taken as the $z$ axis of our coordinate
system. The $x$ axis is horizontal, pointing to the center of the
collider rings and the $y$ axis is vertical, directed upwards.

The EMC barrel and end-caps cover 98\% of the solid angle. Calorimeter modules are read out at both ends by 4880 photomultipliers. Energy and time resolutions are $ \sigma_E /E=~0.057 /\sqrt{E(\rm{GeV})} $ and $ \sigma_t =57\ \rm{ps}/\sqrt{E(\rm{GeV})}\oplus 100\ \rm{ps}$, respectively. 
The drift chamber has only stereo sense wires and is $4$ m in diameter, $ 3.3$ m long. It is built out of carbon-fibers and operates with a low-$Z$ gas mixture (helium with 10\% isobutane). Spatial resolutions are  $\sigma_{xy}\ab150\ \rm\upmu m$ and  $\sigma_z\ab2$ mm. The momentum resolution for large angle tracks is $\sigma(p_\perp) / p_\perp\ab 0.4\% $. The trigger uses both EMC and DC information. Events used in this analysis are triggered by at least two energy deposits larger than 50 MeV in two sectors of the barrel calorimeter~\cite{KLOE_trig}.

\section{Event Selection}\label{Data analysis}

A $\mu \mu \gamma$ candidate must have two tracks of opposite charge,   with the point of closest approach to the $z$ axis within a cylinder of radius 8 cm and length 15 cm centered at the interaction point. We require two tracks to be emitted at large polar angle, $50^\circ\!<\!\theta\!<\!130^\circ$, and an undetected photon whose momentum, computed from the two track values  according to the $\mu \mu \gamma$ kinematics, points at small angle ($\theta\!<\!15^\circ,\ \!>\!165^\circ$)~\cite{KLOE_pi_FF}. These requirements limit the range of $M_{\mu \mu}$ to be larger than 500~MeV.
This separation between the tracks and photon-emission regions greatly reduces the contamination from the resonant process $\epm\to\phi\to\pi^+\pi^-\pi^0$, where charged pions are misidentified as muons and the $\pi^0$ mimics the missing momentum of the photon(s), and from the FSR processes $\epm\to \pi^+\pi^-\gamma_{\,\rm{FSR}}$ and $\epm\to \mu^+\mu^-\gamma_{\,\rm{FSR}}$.
ISR photons are strongly peaked along the beam line.
The above requirements are also satisfied by $\epm\to\epm\gam$ radiative Bhabha events. A particle identification estimator (\Li), based on a pseudo-likelihood function using time-of-flight and calorimeter information (size and shape of the energy deposit) is used to obtain additional separation between electrons and pions or muons~\cite{KLOE1,KLOE2}.

Events with both tracks satisfying $\Li\!<\!0$ are rejected as $\epm\gam$. The signal loss due to this requirement is less than $0.05\%$, as evaluated using $\mu^+ \mu^-$ samples obtained from both measured data and from Monte Carlo (MC) simulation. Pions and muons are identified by means of the variable $M_{\rm trk}$ which is the mass of particles $x^+,\ x^-$ in the $\epm\to x^+ x^- \gam$ process.  We assume the presence of an unobserved photon and that the tracks belong to particles of the same mass and momentum equal to the observed value.
The $M_{\rm trk}$ ranges 80--115 and $>$130 MeV identify muons and pions.

\begin{figure}[htb]
\begin{center}\vspace{-0.2cm}
\includegraphics[width=8cm]{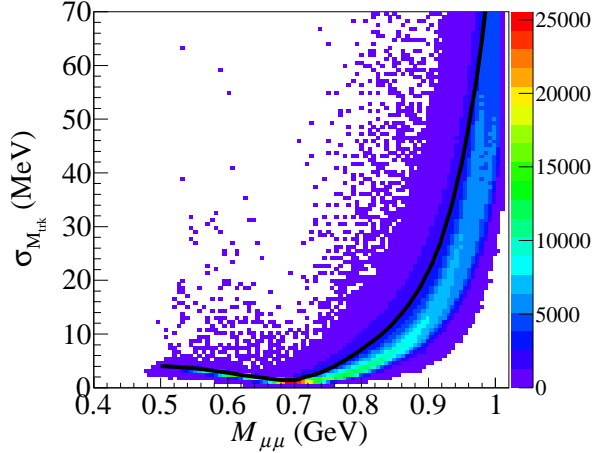}
\caption{Data scatter plot in the $M_{\mu\mu},\ \sigma_{M_{\mathrm{trk}}}$ plane. Events above the solid line are rejected.}
\label{smtrk_cut_new}
\end{center}
\end{figure}
\begin{figure}[htb]
\begin{center}\vspace{-0.2cm}
\includegraphics[width=6.8cm]{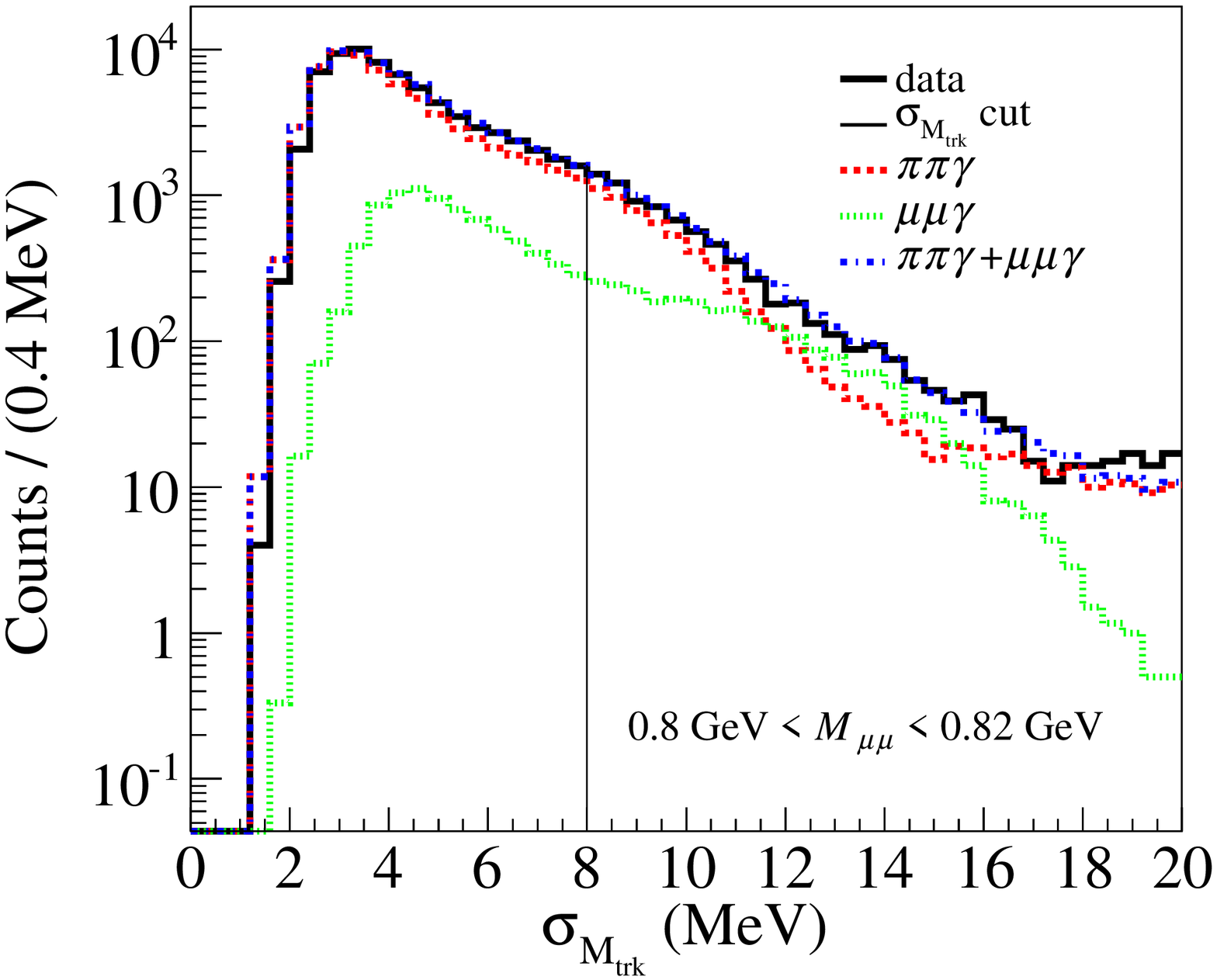}
\includegraphics[width=6.8cm]{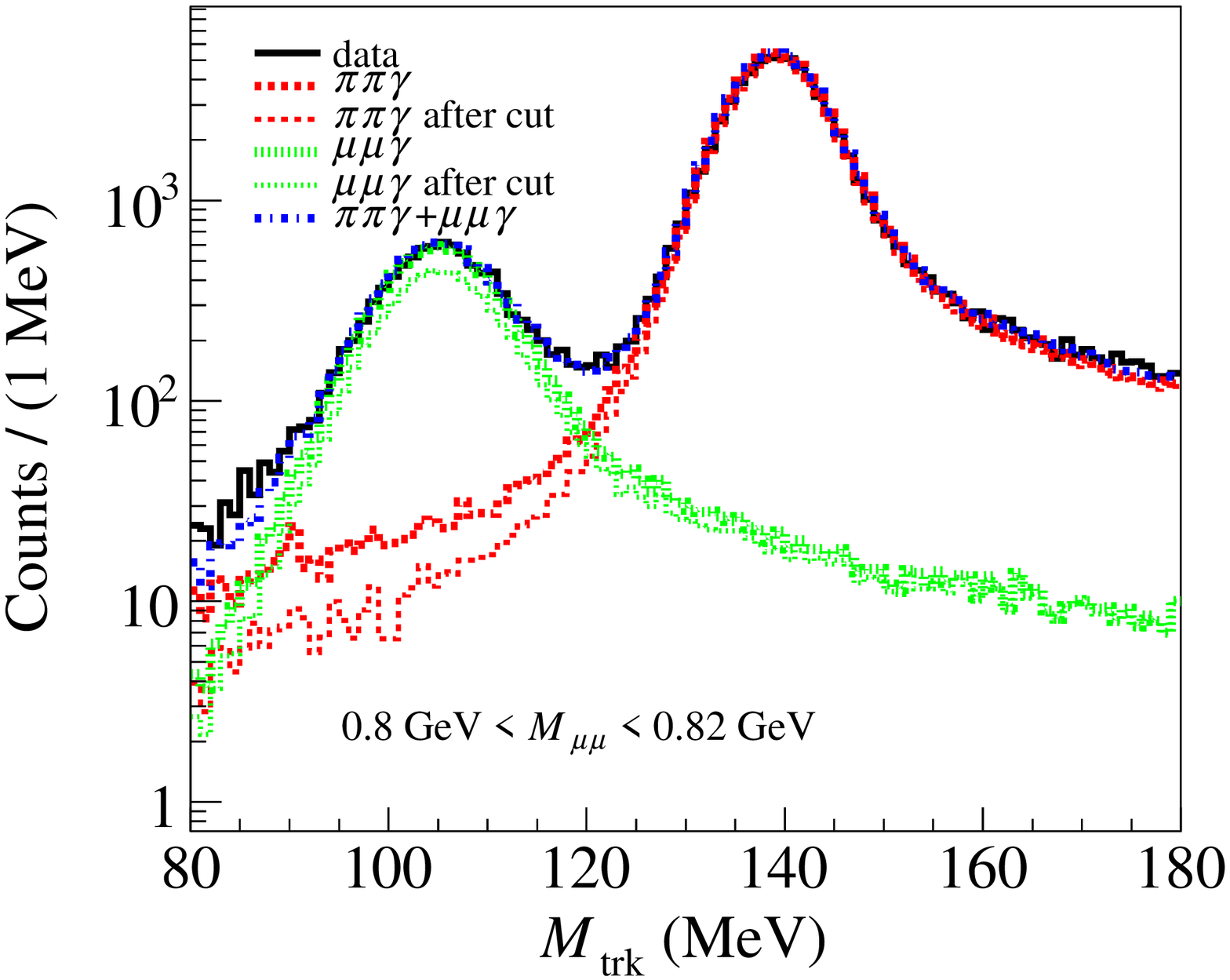}\vspace{-0.2cm}
\caption{
Left: Distribution of $\sigma_{M_{\mathrm{trk}}}$ for one
$M_{\mu\mu}$ slice for data, $\pi^+ \pi^- \gamma$,  $\mu^+\mu^-\gamma$  and the sum of $\pi^+
\pi^- \gamma$ and $\mu^+\mu^-\gamma$. 
The  $\sigma_{M_{\mathrm{trk}}}$ cut is also shown. Right: Effect of the $\sigma_{M_{\mathrm{trk}}}$ cut on \mt\ distributions for
the same slice of $M_{\mu\mu}$ for the \pic\gam ~and $\mu^+\mu^-\gamma,\ M_{\rm{trk}}$ distributions without
$\sigma_{M_{\mathrm{trk}}}$ cut. The corresponding thin lines show the effect of the $\sigma_{M_{\mathrm{trk}}}$ cut. 
All symbols are defined in the figures inserts.}
\label{Sigma_Mtrk}
\end{center}
\end{figure}

 The accuracy of the \mt\ determination depends on the quality of the fitted tracks in the DC.  A variable $\sigma_{M_{\mathrm{trk}}}$, which represents the uncertainty on \mt\ determination, can be constructed. By selecting events with a small $\sigma_{M_{\mathrm{trk}}}$, we were able to create narrower pion and muon peaks in the \mt\ distribution thus improving the $\pi/\mu$ separation.
The $\sigma_{M_{\mathrm{trk}}}$ distribution is correlated with $M_{\mu\mu}$, therefore we apply an  $M_{\mu\mu}$-dependent \mt-cut (whose efficiency varies between 70 and 80\% as function of $M_{\mu\mu}$).
Figure~\ref{smtrk_cut_new}
shows the cut in the $M_{\mu \mu}$, $\sigma_{M_{\mathrm{trk}}}$ plane. The $\sigma_{M_{\mathrm{trk}}}$ distribution  for the slice 0.8 $< M_{\mu\mu}<$ 0.82 GeV is shown in  Fig.~\ref{Sigma_Mtrk} (left). Figure~\ref{Sigma_Mtrk} (right), shows the effect of this cut on the $M_{\rm{trk}}$ distribution, in the same $M_{\mu\mu}$ slice. There is a clear reduction of the left tail of the  \mt\ distribution for $\pi\pi\gam$, resulting in a suppression of the $\pi\pi\gam$  background in the $\mu \mu\gamma$ region, depending at the percent level on the $M_{\mu\mu}$ interval. Figure~\ref{Sigma_Mtrk} also shows a good agreement between data and Monte Carlo simulation  in both $\sigma_{M_{\mathrm{trk}}}$ and  \mt\ variables.

\begin{figure}[htp!]
\begin{center}\vspace{-0.2cm}
\includegraphics[width=8cm]{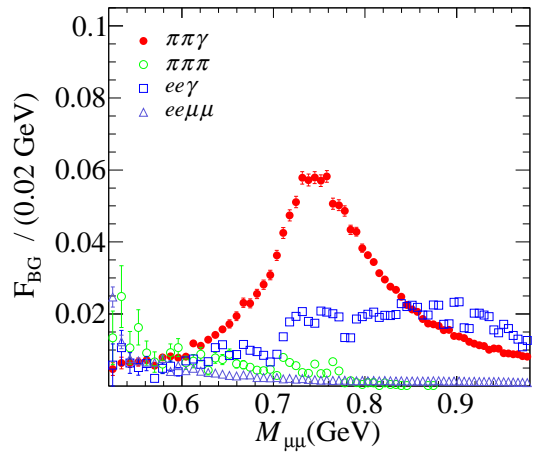}\vspace{-0.2cm}
\caption{ Fractional backgrounds to the $\mu\mu\gam$ signal from the \pic\gam, \pic\po, \epm\gam, and $\epm\mu^+ \mu^-$ channels after all selection criteria, see insert for symbols.}\vspace{-0.2cm}
\label{Bckg}
\end{center}
\end{figure}

 After all the analysis cuts, residual backgrounds consisting of \epm\to\epm\gam, \epm\to\pic\gam\  and  \epm\to\f\to\pic\po\ are still present.
The residual background is obtained by fitting the
observed $M_{\rm trk}$ spectrum  with a superposition of MC simulated distributions describing signal plus \pic\gam, \pic\po\ backgrounds, and a distribution obtained from data for the \epm\gam\ ~\cite{KLOE_pi_FF}.
Additional background from \epm\to\epm$\mu^+\mu^-$ and \epm\to\epm\pic\ has been evaluated.
The \epm\to\epm\pic\ contribution is negligible while  \epm\to\epm$\mu^+\mu^-$ is at the percent level below 0.54 GeV 
and decreases with $M_{\mu\mu}$. 
 Figure~\ref{Bckg} shows the fractions of the background processes, $F_{\rm BG}$, contributing  non-negligibly (only statistical errors are shown), as a function of $M_{\mu\mu}$ after all selection criteria are applied. 
 It's worth noting that no peaking component is seen in the background.

\begin{figure}[htp!]
\begin{center}
\includegraphics[width=6.8cm]{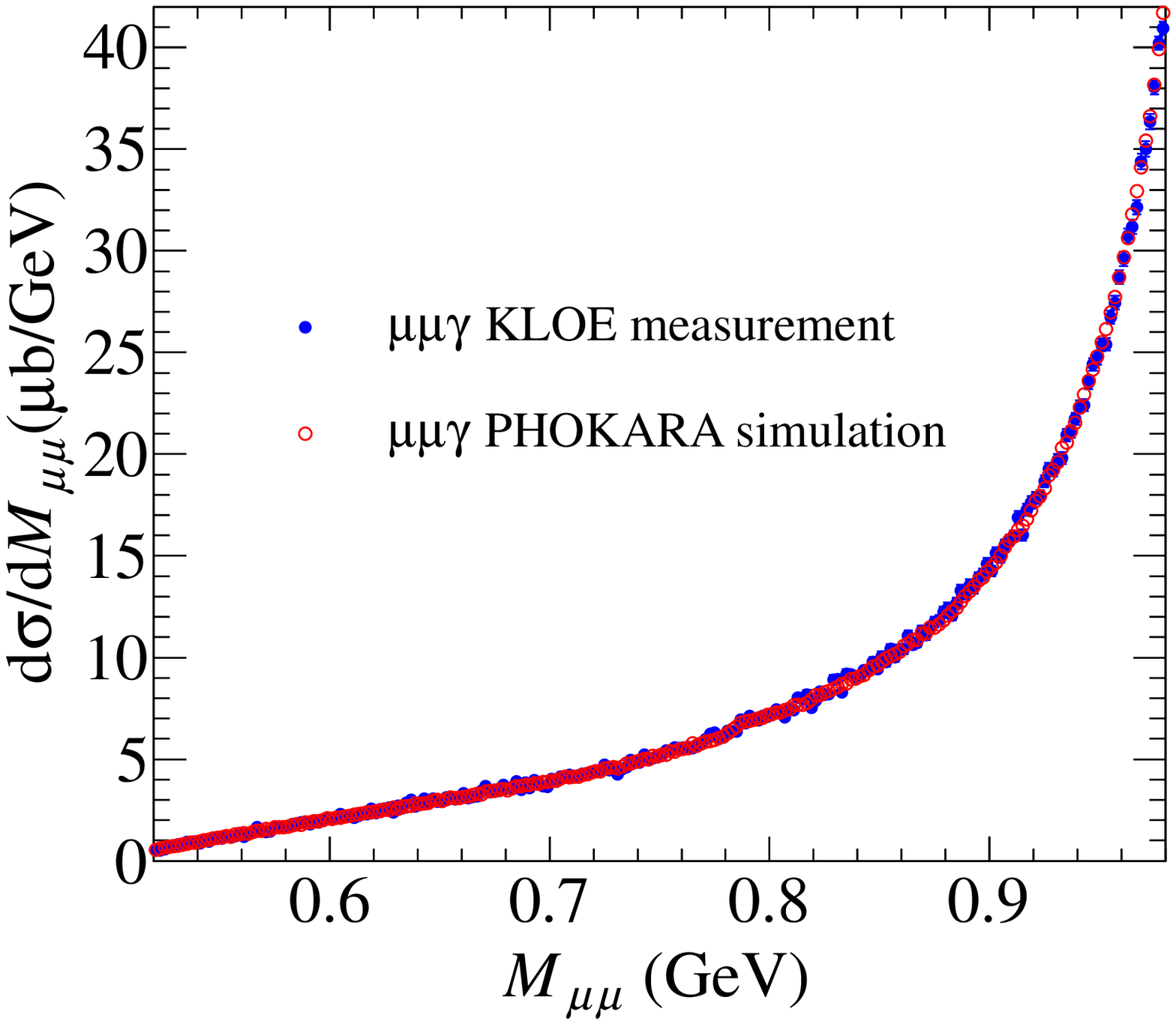}
\includegraphics[width=6.8cm]{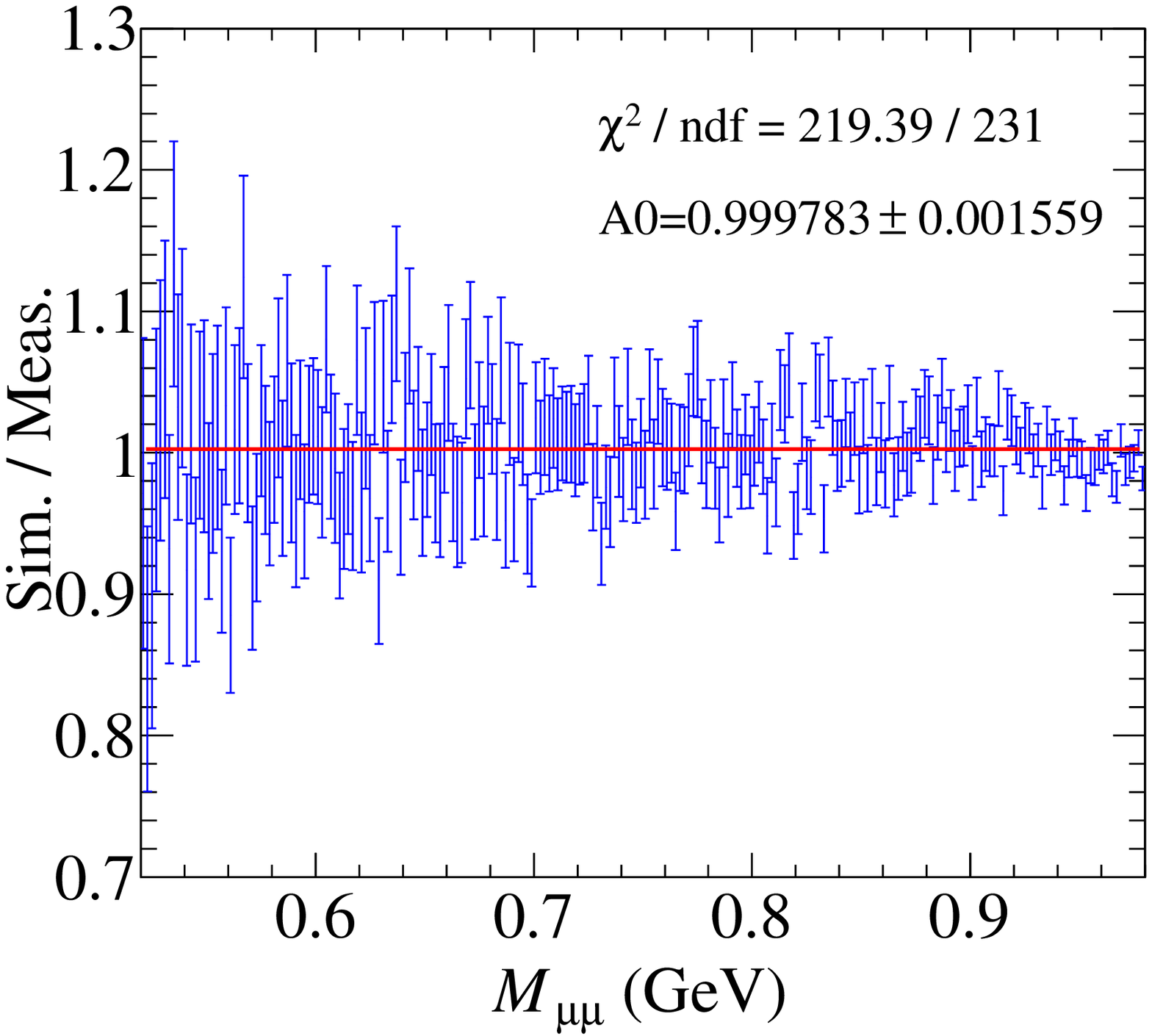}
\caption{Left: Comparison of data (full circles) and simulation (open circles) for $\mu^+ \mu^- \gamma$ cross section. Right: Ratio of the two spectra fitted with a constant function.}
\label{mmg_abs}
\end{center}
\end{figure}

 At the end of  $\mu\mu\gam$  selection criteria, the $M_{\mu \mu}$ spectrum consists  of about 5.35\x10\up5; events.
By correcting it for measurement/simulation difference in tracking and trigger efficiencies (ranging from 0.2 to 1\% as function of $M_{\mu \mu}$), subtracting  the backgrounds and dividing  by the efficiency and integrated luminosity~\cite{PHOKHARA},  we obtain the differential cross section $\dif\sigma_{\mu \mu \gamma}/\dif M_{\mu\mu}$. Figure~\ref{mmg_abs} (left), shows the measured $\mu\mu\gamma$ cross section compared with the NLO QED calculations, using the MC code PHOKHARA~\cite{PHOKHARA}. Figure~\ref{mmg_abs} (right) shows the ratio between the two differential cross sections fitted with a constant function. The  agreement between measurement and the PHOKHARA simulation of the cross section is excellent.

\subsection{Systematic errors and efficiencies}
Several sources of systematic uncertainty contributing to the $\mu^+ \mu^-
\gamma$ event yield estimate have been evaluated.

\textbf{Background subtraction}: the systematic uncertainty is due to the background fit normalization parameters and
the uncertainty on the $\epm\mu^+\mu^-$ residual background.
The total fractional systematic uncertainty, obtained by adding in quadrature the two contributions, ranges from 0.1 to 0.5\%, decreasing with $M_{\mu\mu}$.

\textbf{$M_{\rm \bf trk}$ cut}: the $M_{\rm trk}$ selection region for $\mu^+ \mu^- \gamma$  is $80<M_{\rm trk}<115$~MeV. We varied the region boundaries by 5 MeV, and computed the ratio of the measured cross sections 
in the new region and in the region of standard cuts.
The systematic uncertainty (constant in $M_{\mu \mu}$) is 0.4\%.

{\bf\mathversion{bold}$\sigma_{M_{\mathrm{trk}}}$ cut}: the systematic uncertainty has been evaluated as the maximum difference between the $\mu \mu \gamma$ normalization parameters of the background fitting procedure, obtained with standard cuts, and those obtained by shifting $\sigma_{M_{\mathrm{trk}}}$ by $\pm$5\%.
The systematic contribution reaches the percent level (up to a maximum of 1.2\%) at low $M_{\mu \mu}$  and decreases below 1\% for $M_{\mu\mu} > 0.76$~GeV.

\textbf{Acceptance}: we estimate the uncertainty resulting from the angular acceptance cut for muons and photon to range from 0.1 up to 0.6\%, by varying the limits by 1\deg.

\textbf{Tracking}: the single muon tracking efficiency, as function of the particle momentum and polar angle is obtained by a high purity $\mu^+ \mu^- \gamma$ sample  using one muon to tag the presence of the other.
The combined efficiency is about 99\%, almost constant in $M_{\mu\mu}$. The systematic uncertainty on tracking efficiency is evaluated changing the purity of the control sample and ranges from $0.3$ to $0.6\%$ as a function of $M_{\mu \mu}$.

\textbf{Trigger}: the trigger efficiency has been obtained from a sample of $\mu^+\mu^-\gamma$ events where a single muon satisfies the trigger requirement. Trigger response for the other muon is parameterized as a function of its momentum and direction. The efficiency as a function of $M_{\mu\mu}$  is obtained using the MC event distribution and differs from one by less than $ 10^{-3}$ for $M_{\mu\mu}<0.6$~GeV and less than $10^{-4}$ for $M_{\mu\mu}>0.6$~GeV.

\textbf{Radiator function}:  we take as systematic uncertainty on $H$ the value of 0.5\%, as quoted in Refs.~\cite{H,H_1,H_2,PHOKHARA}.

\textbf{Luminosity}: we calculated the luminosity using large-angle Bhabha scattering events~\cite{PHOKHARA}, and evaluated the related systematic uncertainty to be 0.3\%. 

\begin{figure}[htp!] \begin{center} \vspace{-0.3cm}
\includegraphics[width=6.7cm]{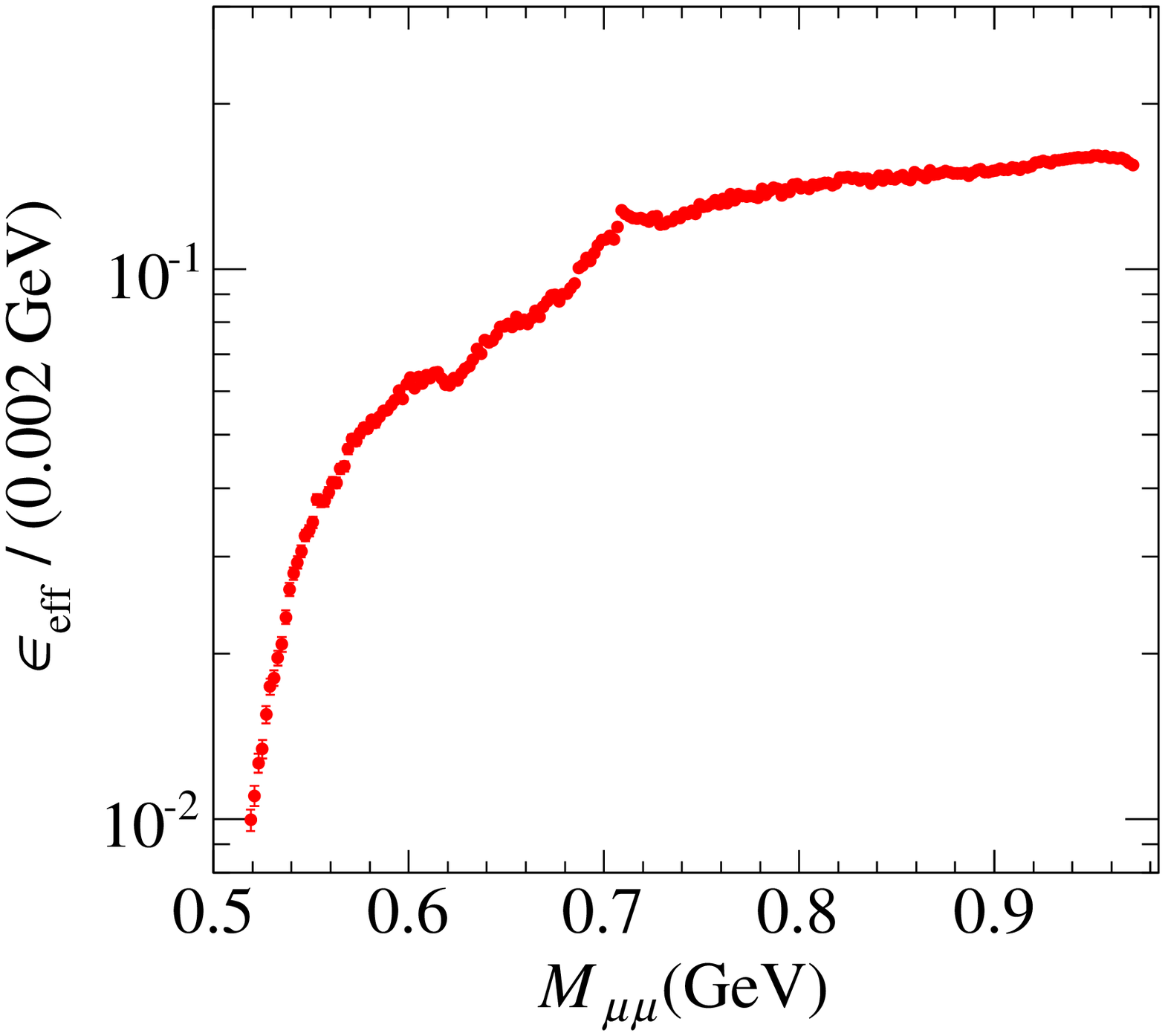}
\includegraphics[width=6.7cm]{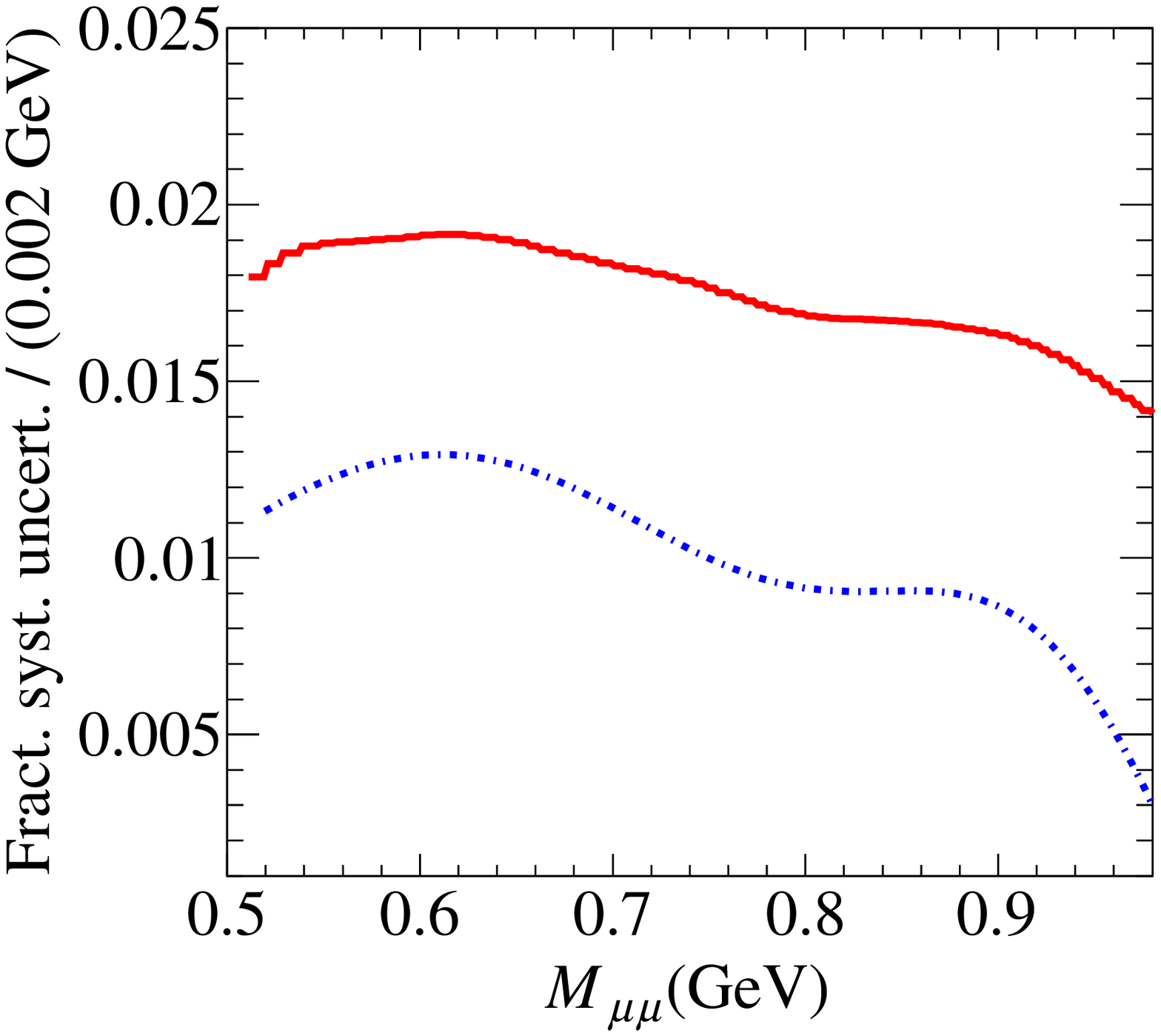}
\caption{Left: Global analysis efficiency. Right: Total fractional systematic uncertainty on the expected $\mu \mu \gamma$ yield (solid line), the contribution  due to the $\sigma_{M_{\rm trk}}$ cut only  is shown by the dash-dotted line.}  
\label{tot_tlimit_syst} \end{center}\vspace{-0.4cm} \end{figure}

The dependence of the reconstruction efficiency, $\epsilon_{\rm eff}$, on $M_{\mu\mu}$ is shown in Fig.~\ref{tot_tlimit_syst} (left). In Fig.~\ref{tot_tlimit_syst} (right), we show the dependence of the total systematic uncertainty on the $\mu^+\mu^-\gamma$ yield.

The largest contribution to total systematic error comes from
the uncertainty on $\sigma_{M_{\mathrm{trk}}}$ cut, as shown by the dash dotted line in the figure.

\section{\mathversion{bold}Upper Limit on $U$-boson Coupling}
\label{UL}

$U$-boson  decays into $\mu^+ \mu^-$ would appear as a peak over the smooth $\mu^+ \mu^- \gamma$ QED contribution. We extract the limit on the number of $U$-boson candidates
by using the CLS technique~\cite{CLS_Technique,Junk,Read_cls}.
As \textit{data} input for the limit extraction procedure, we use the
observed invariant mass distribution. As \textit{background} input,  we used the
$\mu^+\mu^-\gamma$ events simulated with PHOKHARA with the addition of the  background sources reviewed in Section~\ref{Data analysis}.
The $U$-boson signal is 
generated, for each $M_{U}$ value, through a toy MC with a gaussian shape.
The signal width takes into account the resolution in $M_{\mu\mu}$ which varies from 1.5~MeV to 1.8~MeV, as $M_{\mu\mu}$ increases.
The mass resolution has been checked by comparing momenta, track mass, and the error of the track mass distributions of data and MC simulation.
The related uncertainty on the $U$-boson mass shape is negligible due to the bin width (2 MeV).
\begin{figure}[htp!]

\begin{center}
\includegraphics[width=9cm]{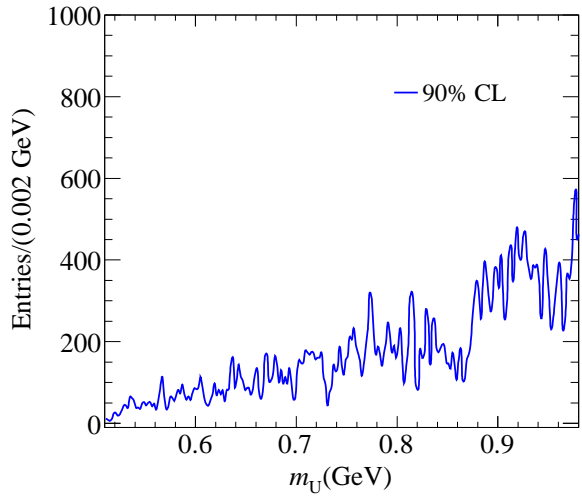}\vspace{-0.2cm}
\caption{Upper limit on the number of signal events at the $90\%$ CL as function of the $U$-boson mass $M_{\rm U}$.}
\label{Raw_TLimit_Output}
\end{center}
\end{figure}

Figure~\ref{Raw_TLimit_Output} shows the upper limit on the number of signal events ($N_{\rm CLS}$) at 90\% confidence level (CL), computed in steps of 2 MeV. A total systematic uncertainty between 1.4 and 1.8\%, as shown in Fig.~\ref{tot_tlimit_syst}, has been applied to the background. We find no evidence for a signal and we therefore set an upper limit on the kinetic mixing parameter $\epsilon^2$ at 90\% CL. 
We extract the limit on the kinetic mixing parameter according to
\begin{equation}
 \epsilon^2= \frac{N_{\rm{CLS}
  }/ (\epsilon_{\rm{eff}} \cdot L)}{H \cdot I},
\label{eq.3}
\end{equation}
where  $\epsilon_{\rm{eff}}$ represents
the overall efficiency, $L$ is the integrated luminosity, $H$ is the radiator function and $I$ is the effective cross section~\cite{Fayet} for $\epm\to U\to \mu^+\mu^-$ integrated on a single mass bin with $\epsilon=1$.  
The $U \to \mu^+ \mu^-$ branching fraction uncertainty ranges from 0.5\% at 500 MeV to 2\% at the
$\rho-\omega$ peak and has been included in the UL extraction on $\epsilon^2$.
In the present analysis we assume that the $U$ boson decays  into SM particles only.

\begin{figure}[h]
\begin{center}
\includegraphics[width=8cm]{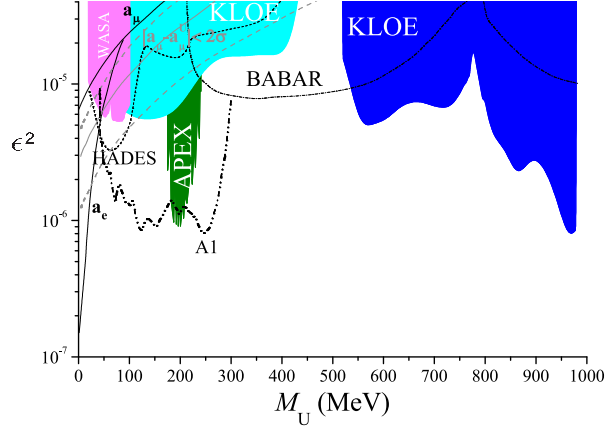}
\caption{90\% CL exclusion plot  for $\epsilon^2$ as a function of the $U$-boson mass (blue). The limits from  the A1~\cite{Mami, Mami1} (dashed double dotted) and Apex~\cite{Apex} (green) fixed-target experiments, the KLOE limit from $\phi\to\epm\gam$~\cite{KLOE_UL2} (cyan), the WASA~\cite{WASA} (magenta) and  HADES~\cite{HADES} limits (dashed line) are also shown. The dash-dotted line is an estimate using BaBar data~\cite{Reece,BaBar}. The solid lines are the limits from the muon and electron anomaly~\cite{a_mu}, respectively. The gray line shows the U-boson parameters that could explain the observed $a_{\mu}$ discrepancy with a $2 \sigma$ error band (gray-dashed lines)~\cite{a_mu}.}
\label{Exclusion_plot}
\end{center}
\end{figure}

The resulting exclusion plot on the kinetic mixing parameter
$\epsilon^2$, in the 520--980~MeV mass range,  is shown in
Fig.~\ref{Exclusion_plot}.
The sensitivity loss due to the $\rho$ meson around 770~MeV is  visible.
In the same plot,  other limits in the mass range below
$1$ GeV are also shown~\cite{Mami,Mami1,Apex,WASA,KLOE_UL1, KLOE_UL2,Reece,BaBar}.
The solid black lines are the limits from the muon and electron $g-2$~\cite{a_mu}. The gray line shows the U-boson parameters that could explain the observed $a_{\mu}$ discrepancy with a $2 \sigma$ error band (gray-dashed lines)~\cite{a_mu}.  Our 90\% CL limit is between 1.6$\x$10$^{-5}$ and 8.6$\x$10$^{-7}$ in the
520--980~MeV mass range.

\section{Conclusions}
\label{conclusions}

We have  searched  for a light, dark vector boson in the $\epm\to\mu^+\mu^-\gamma$ channel in a sample of 5.35\x10\up5; events recorded with the KLOE detector for a total integrated luminosity of 239.3~pb$^{-1}$. We find no evidence for a $U$ boson in the mass range 520--980~MeV. We set an upper limit at 90\% CL on the kinetic mixing parameter $\epsilon^2$ between  \pt1.6;-5; and \pt8.6;-7;. 
The limit is derived through a study of the $\mu^+\mu^-\gamma$ ISR process and significantly improves the current limit on $\epsilon$ in this mass range.

\section{Acknowledgments}
We warmly thank our former KLOE colleagues for the access to the data collected during the KLOE data taking campaign.
We thank the \DAF\ team for their efforts in maintaining low background running conditions and their collaboration during all data taking. We wish to thank our technical staff:
G.F. Fortugno and F. Sborzacchi for their dedication in ensuring efficient operation of the KLOE computing facilities;
M. Anelli for his continuous attention to the gas system and detector safety;
A. Balla, M. Gatta, G. Corradi and G. Papalino for electronics maintenance;
M. Santoni, G. Paoluzzi and R. Rosellini for general detector support;
C. Piscitelli for his help during major maintenance periods.
This work was supported in part by the EU Integrated Infrastructure Initiative Hadron Physics Project under contract number RII3-CT- 2004-506078; by the European Commission under the 7th Framework Programme through the `Research Infrastructures' action of the `Capacities' Programme, Call: FP7-INFRASTRUCTURES-2008-1, Grant Agreement No. 227431; by the Polish National Science Centre through the Grant Nos. 0469/B/H03/2009/37, 0309/B/H03/2011/40, DEC-2011/03/N/ST2/02641, \\
2011/01/D/ST2/00748, 2011/03/N/ST2/02652, 2013/08/M/ST2/00323 and by the Foundation For Polish Science through the MPD programme and the project HOMING PLUS BIS/2011-4/3.

\end{document}